\documentclass[default,iicol]{sn-jnl}

\usepackage{graphicx}%
\usepackage{multirow}%
\usepackage{amsmath,amssymb,amsfonts}%
\usepackage{amsthm}%
\usepackage{mathrsfs}%
\usepackage[title]{appendix}%
\usepackage{xcolor}%
\usepackage{textcomp}%
\usepackage{manyfoot}%
\usepackage{booktabs}%
\usepackage{multicol} 
\usepackage{subfig}
\usepackage{algorithm}%
\usepackage{algorithmicx}%
\usepackage{algpseudocode}%
\usepackage{listings}%

\theoremstyle{thmstyleone}%
\theoremstyle{thmstyletwo}%

\theoremstyle{thmstylethree}%

\begin{document}

\title[Article Title]{Efficient Superconductor Arithmetic Logic Unit for Ultra-Fast Computing}


\author*[1,2]{\fnm{Sasan} \sur{Razmkhah}}\email{razmkhah@usc.edu}

\author[1]{\fnm{Ali} \sur{Bozbey}}\email{bozbey@etu.edu.tr}
\equalcont{These authors contributed equally to this work.}

\affil[1]{\orgdiv{Department of Electrical and Electronics Engineering}, \orgname{TOBB University of Economics and Technology}, \orgaddress{\street{Sogutozu}, \city{Ankara}, \postcode{06560}, \country{Turkey}}}

\affil[2]{\orgdiv{Current Affiliation: Ming Hsieh Department of Electrical and Computer Engineering}, \orgname{University of Southern California}, \orgaddress{\street{3740 McClintock}, \city{Los Angeles}, \postcode{90089}, \state{CA}, \country{USA}}}


\abstract{We present a 4-bit Arithmetic Logic Unit (ALU) utilizing superconductor technology. The ALU serves as the central processing unit of a processor, performing crucial arithmetic and logical operations. We have adopted a bit-parallel architecture to ensure an efficient and streamlined design with minimal fanin/fanout and optimal latency. In terms of fabrication, the ALU has been fabricated using a standard commercial process. It operates at an impressive clock frequency exceeding 30 GHz while consuming a mere 4.75 mW of power, including applied reverse current, encompassing static and dynamic components. The ALU contains over 9000 Josephson junctions, with approximately 7000 JJs dedicated to wiring, delay lines, and path balancing, and it has over 18\% bias margin. Designed as a co-processor, this arithmetic logic unit will work with external CMOS memory and processors via interface circuits. Thorough testing and validation of the ALU's functionality have been conducted with digital and analog simulations, and all the components were fabricated and measured within a 4~K pulse-tube cryocooler. Experimental verification has confirmed the successful operation of both the arithmetic and logic units. These results have been analyzed and are presented alongside the experimental data to provide comprehensive insights into the ALU's behavior and capabilities.}




\maketitle

\section{Introduction}\label{sec1}

In the pursuit of high-speed and low-power computing, there is a growing demand for alternative logic families to complement CMOS technology. Recent advancements in superconductor logic technology and superconducting very large-scale integration (VLSI) circuit fabrication have enabled the design of complex rapid single flux quantum (RSFQ) circuits and structures with a significant number of Josephson junctions integrated onto a single chip \cite{ref1,ref2}. These breakthroughs have paved the way for the development of logic circuits that consume orders of magnitude less power than CMOS and operate at considerably higher frequencies \cite{ref3}.

Power consumption poses a critical challenge in large-scale computer systems, contributing to carbon emissions and sustainability concerns \cite{ref4}. The issue also extends to CMOS interconnections and cooling on a smaller scale. With the growth of internet traffic, AI model training, and cloud computing, data center facilities are expanding rapidly. The reliance on cloud computing and artificial intelligence in cellular and mobile devices further amplifies the need for support \cite{ref5, ref6}. The Department of Energy (DoE) and the Defense Advanced Research Projects Agency (DARPA) are actively working on optimizing and minimizing power consumption in these systems \cite{ref7}. Since 2010, however, the data center landscape has changed dramatically due to more efficient computing. However, by 2018, global data center workloads and compute instances had increased more than 6$\times$, whereas data center internet protocol (IP) traffic had increased by more than 10$\times$ \cite{ref8}. This has a significant environmental impact and raises whether computing efficiency can compete with the increasing data demand \cite{ref9}. However, limited information is publicly available due to the private nature of these facilities.

Energy consumption in logic devices arises from various sources, including resistors used for circuit biasing, energy dissipation in interconnects, and bit loss during circuit switching. The Shannon-Neuman-Landauer limit \cite{ref10} defines the theoretical threshold for energy dissipation in logic gates due to bit loss during switching. Superconductor technology offers the potential to approach power consumption near this theoretical limit.

Over the past few decades, researchers have dedicated significant efforts to developing stable superconducting processors and co-processors \cite{ref11,ref12,ref13}. Among the early endeavors, Semenov et al. embarked on designing arithmetic logic units (ALUs) using RSFQ logic in the 1990s \cite{ref14}. Their work focused on a bit-serial architecture with approximately 5000 junctions and aimed to achieve a clock frequency of up to 20 GHz. Another notable milestone in RSFQ processor implementation was the FLUX project by Bunyk et al. at SUNY Stony Brook in 2001 \cite{ref11}. They aimed to fabricate a 16-bit ALU incorporating registers and an instruction set memory supporting over 30 commands. The chip they designed contained over 90000 junctions and occupied an area of $10\times15 mm^2$. Building upon this, Bunyk et al. continued their research on ALU architecture and developed the FLUX-1 microprocessor in 2003 \cite{ref15}. The Flux-1 chip was an 8-bit microprocessor designed for a frequency of 20 GHz, featuring a parallel partitioned architecture encompassing 63000 Josephson junctions within a $10\times10 mm^2$ area. In 2004, Kang et al. from Incheon University in South Korea designed an RSFQ ALU employing half adders in a serial pipeline architecture \cite{ref16}. In 2006, Müller et al. from Stellenbosch University in South Africa developed a superconductor ALU using RSFQ-AT gates \cite{ref17}.

One of the notable ongoing efforts in the field is led by Fujimaki et al. at Nagoya University in Japan. In 2015, they successfully designed a 4-bit bit-slice Arithmetic Logic Unit (ALU) integrated with a 32-bit RSFQ microprocessor \cite{ref13}. To fabricate the circuits, they employed the advanced process (ADP), a nine-layer Nb process with a current density of 10 $kA/cm^2$. The chip had dimensions of 3$\times$1.7 $mm^2$ and operated at a clock frequency of 50 GHz. Fujimaki's laboratory group at Nagoya University has also developed the CORE-e series of RSFQ-based microprocessors, which are general-purpose microprocessors with random access memories and bit-serial architecture \cite{ref18}. They continued their work and developed an 8-bit processor with 100TOPs/W in \cite{ref19}.

However, all these circuits were tested in liquid helium cryostats, which are not commercially viable for widespread application due to high operational costs and limited operation time. In this work, we aimed to design an Arithmetic Logic Unit (ALU) utilizing RSFQ logic to test it in a closed-cycle cryostat system and demonstrate the capability of the technology for commercial use. We use a parallel bit propagation data path. The parallel architecture, although relatively complex compared to the serial ALU, offers a significantly higher speed at the expense of a higher bias current. The architecture is pipelined, and new data and instruction sets can applied in each clock cycle. In this study, we provide a comprehensive explanation of the ALU's structure, followed by circuit-level design, digital and analog simulations, and the layout design of the circuit. Additionally, we present the experimental test setup and measurement results for the individual arithmetic and logic units, providing valuable insights into their performance and functionality.

\section{Architecture}\label{sec2}

A processor requires an Arithmetic Logic Unit (ALU), reliable and dense cache memory, and efficient data buses. However, a reliable dense memory design remains challenging in RSFQ technology \cite{ref20,ref21}. One potential solution is incorporating a superconductor ALU as a co-processor alongside a CMOS processor. Integrating a CMOS processor with an RSFQ ALU can eliminate the need for extensive memory at the superconductor stage. A relatively small register memory would suffice for storing commands and in-line data, while the primary memory resides on the CMOS circuit. The CMOS circuits could be positioned in the first stage of the cryocooler at 55K, leveraging the benefits of both technologies.

It is important to note that the RSFQ clock operates at a significantly higher frequency than the CMOS clock. Therefore, input and output stages are required to buffer the data and enable seamless communication between the RSFQ and CMOS processors without compromising computational speed or data integrity. These buffer registers would store the data and instruction sets and synchronize the clocks of both CMOS and RSFQ circuits. The input buffer of the input stage would store the data using the incoming CMOS clock and transfer it to the RSFQ ALU utilizing a series of fast clock pulses precisely timed on the chip. This ensures efficient data transfer and coordination between the two processing units while harnessing the advantages of their respective clock frequencies.
\begin{figure}[htbp]
  \includegraphics[width=0.95\linewidth]{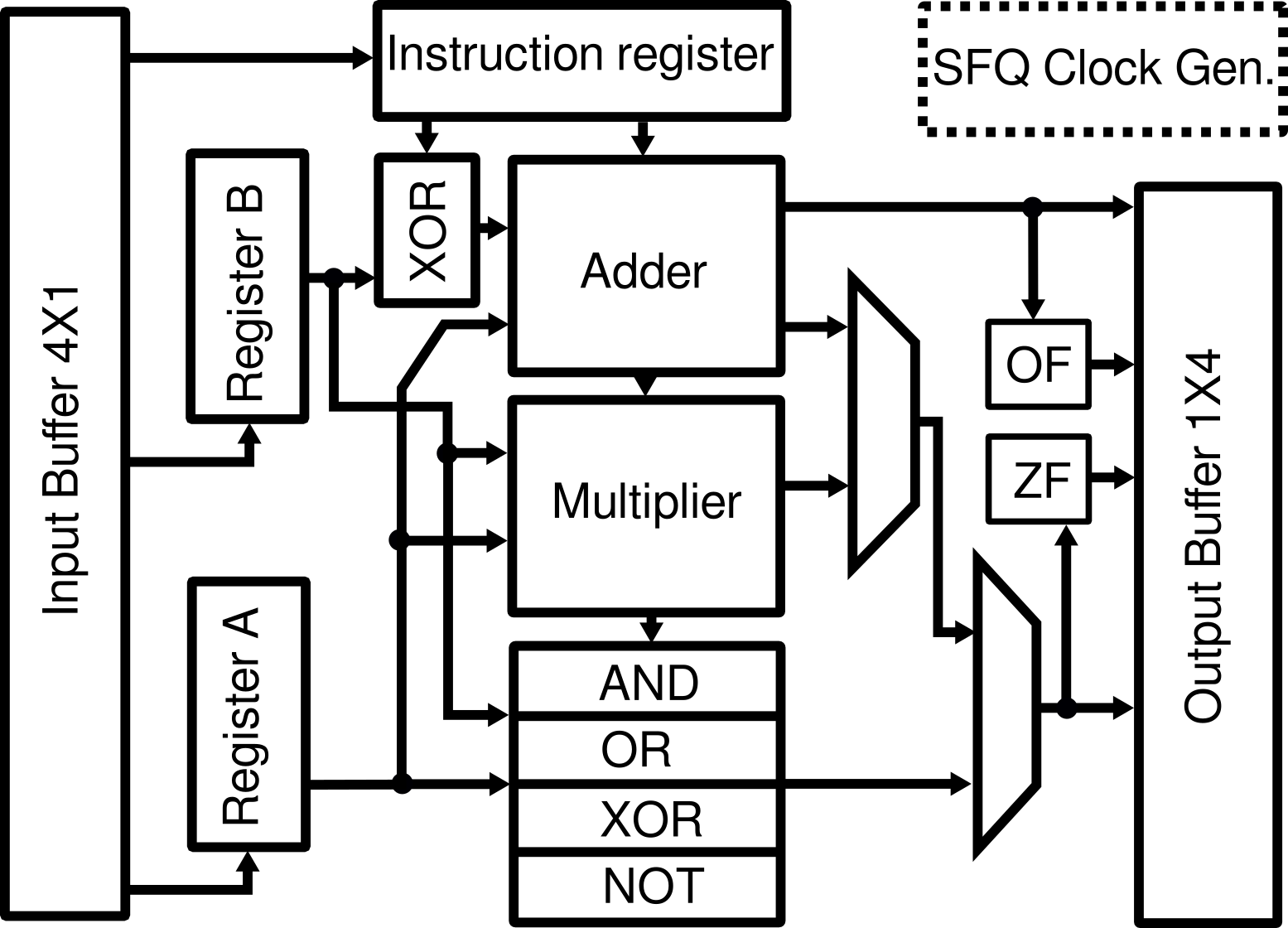}
  \caption{A simplified block diagram of parallel ALU showing the connection between arithmetic and logical operators. The fast and slow clock distribution is not shown here.}%
  \label{figure1} 
\end{figure}

\begin{table}[h]
\caption{Parallel ALU operations and the select bits for each operator.}\label{table2}%
\begin{tabular}{@{}llll@{}}
\toprule
Operation & $S_0$ & $S_1$ & $S_2$\\
\midrule
XOR ($A\oplus B$)& 10 & 10 & 11 \\
OR ($A|B$)&  11  & 11 & 11 \\
AND ($A\&B$)& 10 & 11 & 11 \\
NOT ($~A$)& 11 & 10 & 11 \\
ADD ($A+B$)& 00 & 10 & 10 \\
SUB ($A-B$)& 10 & 10 & 10 \\
MUL ($A\times B$)& XX & 11 & 10 \\
\botrule
\end{tabular}
\end{table}

When designing an Arithmetic Logic Unit (ALU), there are various architectural choices available for the arithmetic unit. The architecture for each unit is selected based on the number of levels and complexity. Adders can be serial, parallel, bit-sliced, and Kogge-Stone architectures \cite{ref12,ref22,ref23,ref24} while multiplier can be serial, parallel, array, Wallace tree, Booth, and many more. Each architecture has its advantages and disadvantages. SFQ gates have a fanout of one, and they rely on clocked operations and can be pipelined for high throughput. As the architecture becomes more complex, the clock tree consumes significant power and introduces delays that drop the clock frequency. The Kogge-Stone architecture offers suitable delay and latency characteristics. However, as demonstrated in our previous work, its complex clock tree and high fanout result in increased power consumption and a larger circuit footprint \cite{ref24}. Although less complicated than Kogge-Stone, the ripple carry adder exhibits higher output latency when scaled. On the other hand, while the serial architecture is easily scalable, it becomes less practical for higher bit numbers, and the output delay and data path imbalance can limit the benefits of the fast clock offered by RSFQ technology. Nonetheless, many RSFQ logic units have been designed using the serial architecture due to its ease of implementation and the ability to increase the number of bits simply by cascading the units \cite{ref12,ref13,ref22,ref25-1,ref26-1}.

For our specific design, we have opted for the parallel datapath for the ALU with a ripple carry adder and array multiplier. Fig.~\ref{figure1} illustrates the block diagram of our chosen method. The components of this ALU include a 4-bit multiplier, 4-bit logic gates (And, Or, Xor, and inverter gates), adder and subtractor circuits, and three flag bits. Overall, the ALU comprises eight direct instruction sets. The operators and their corresponding instruction sets are listed in Table~\ref{table2}. To facilitate seamless integration with CMOS, the ALU generates three flag signals: Zero, Negative, and Carry. The clock output pin determines the critical path of the signal and provides the handshake for the interface circuits. Input and output registers have been designed and fabricated to interface with CMOS and synchronize the inputs, instructions, and outputs, ensuring efficient data flow and coordination. The interface circuits will buffer data and match the low-speed CMOS clock to the high-speed SFQ clock.
\section{Results and Discussion}\label{Sec3}

Before fabrication, we simulated all sub-circuits to ensure their correct functionality. This pre-fabrication simulation allowed us to verify the performance of each sub-part within the ALU design. Subsequently, these sub-parts were fabricated independently, and we tested them in our measurement setup. This rigorous testing process ensured that each component of the ALU met the desired specifications and performed as intended.
\subsection{Simulation Results}
\begin{figure}[htpb]
  \subfloat[Verilog simulation]{\includegraphics[width=0.8\linewidth]{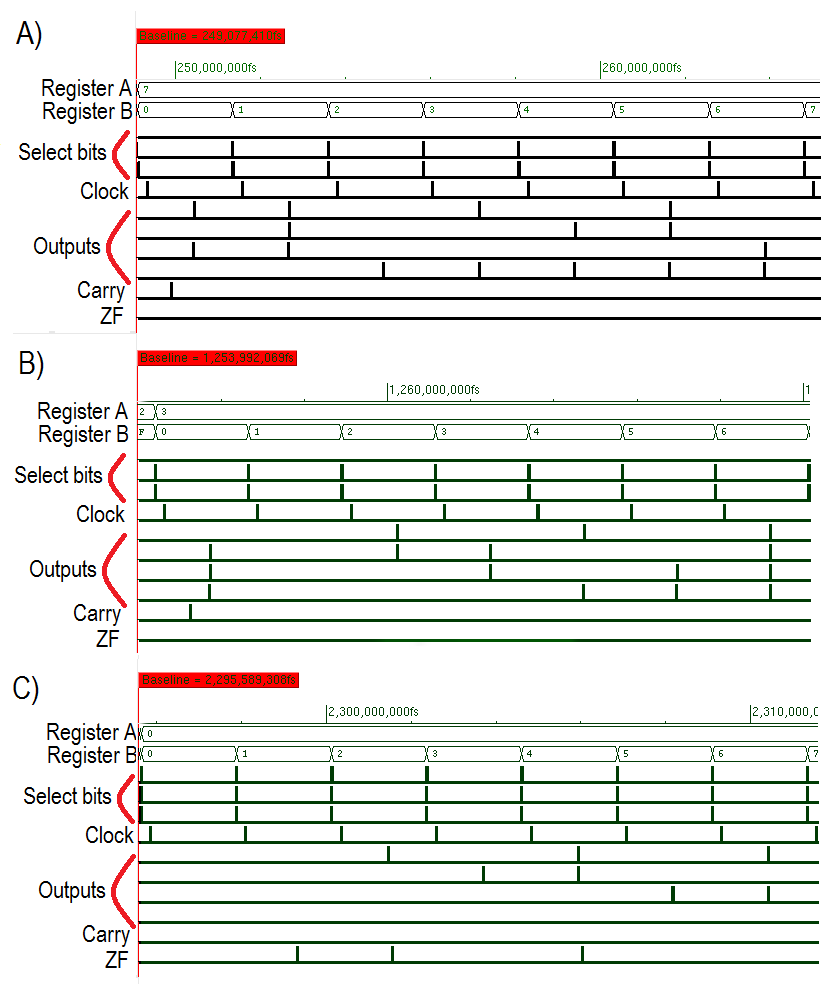}}
  \hfill
  \subfloat[Jsim simulation]{\includegraphics[width=0.8\linewidth]{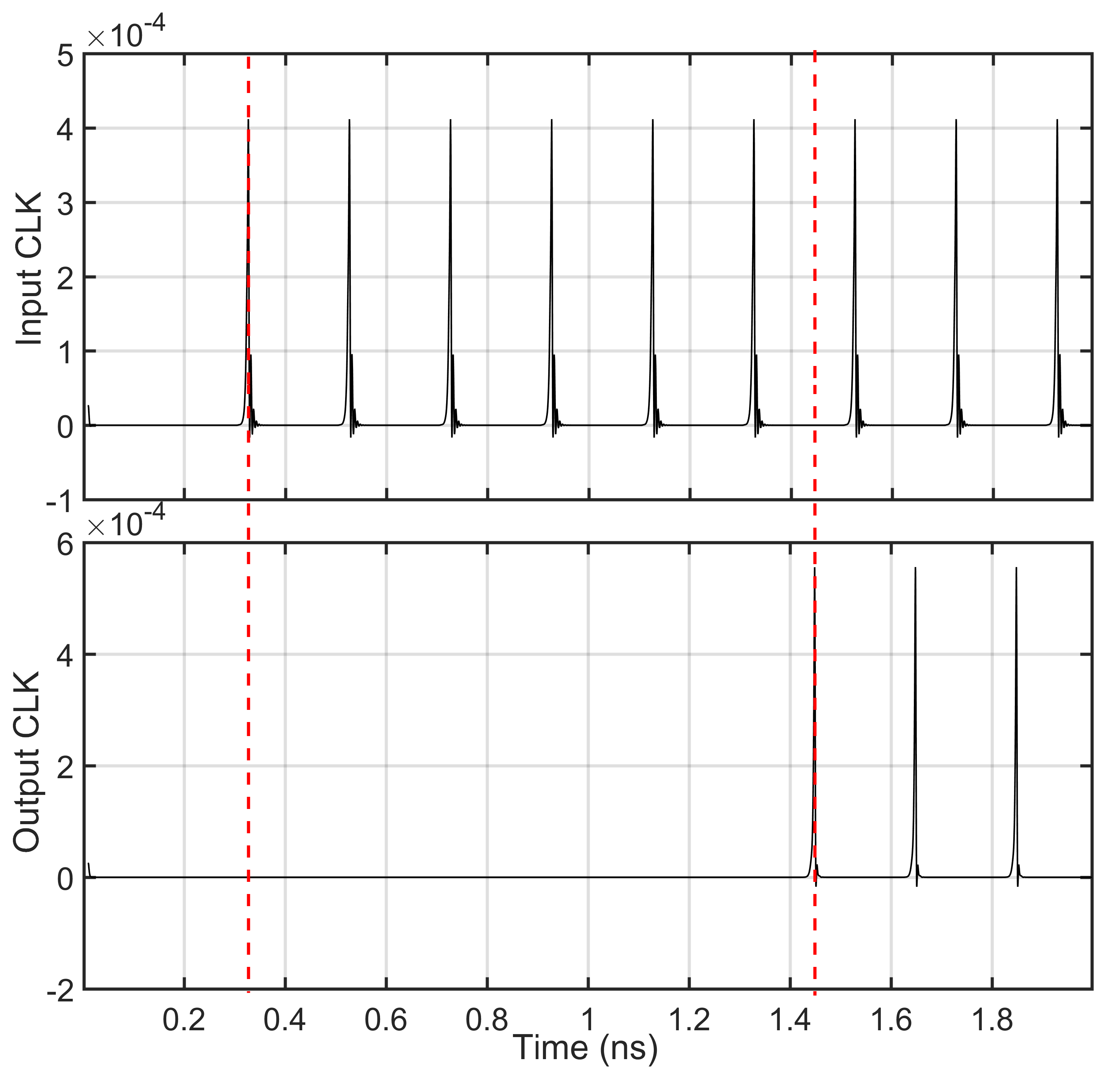}}
  \caption{ALU digital (a) and analog (b) simulation results. (a) Verilog simulation for the ALU was run on all the possible input patterns, and the results were verified. A) The result for the adder as the inputs change. B) Result for the multiplier. C) Result for the Josephson OR gate. (b) The JSIM analog simulation of the ALU circuit on its most critical path (The clock tree).}
  \label{figure2ab}
\end{figure}

The units and all the interface circuits were tested individually with precise cell-based verilog models. then the ALU underwent digital simulation using Verilog, with various bias values employed to assess the latency and operational characteristics for all possible input combinations. The digital simulation results for the ALU, as implemented in Verilog, are presented in Fig.~\ref{figure2ab}(a). This figure illustrates the application of all possible input combinations to the circuit, with the resulting outputs analyzed and evaluated. To examine the ALU's performance under different bias conditions, the bias was swept within a $\mathrm{\pm25\%}$ around the design value of 2.5 mV, ensuring a comprehensive investigation of the ALU's functionality across a range of biases and operation in $\mathrm{\pm18\%}$ margin was observed.
While digital simulation is faster, analog circuit-level simulation will give us more precise results. To conduct analog simulation, we utilized JSIM, a Spice-based simulator designed explicitly for Josephson junction-based circuits \cite{ref27-1}. The simulation focused on monitoring the output clock signal along the most critical path within the chip, considering various bias values. Fig.~\ref{figure2ab}(b) illustrates the logic unit's input and output clock signals along this critical path. By simulating the latency of the output signal on this path, we determined that the expected latency under design conditions is approximately one nanosecond. The analog simulation provided valuable insights into the performance and timing characteristics of the ALU design. However, due to the scale and complexity of the circuit, unlike Verilog simulation for analog, a few selected patterns on the most critical path were tested.

\subsection{Fabricated Circuits}
\begin{figure}[htbp]
 \subfloat[Schematic of a 4-bit multiplier]{\includegraphics[width=0.5\linewidth]{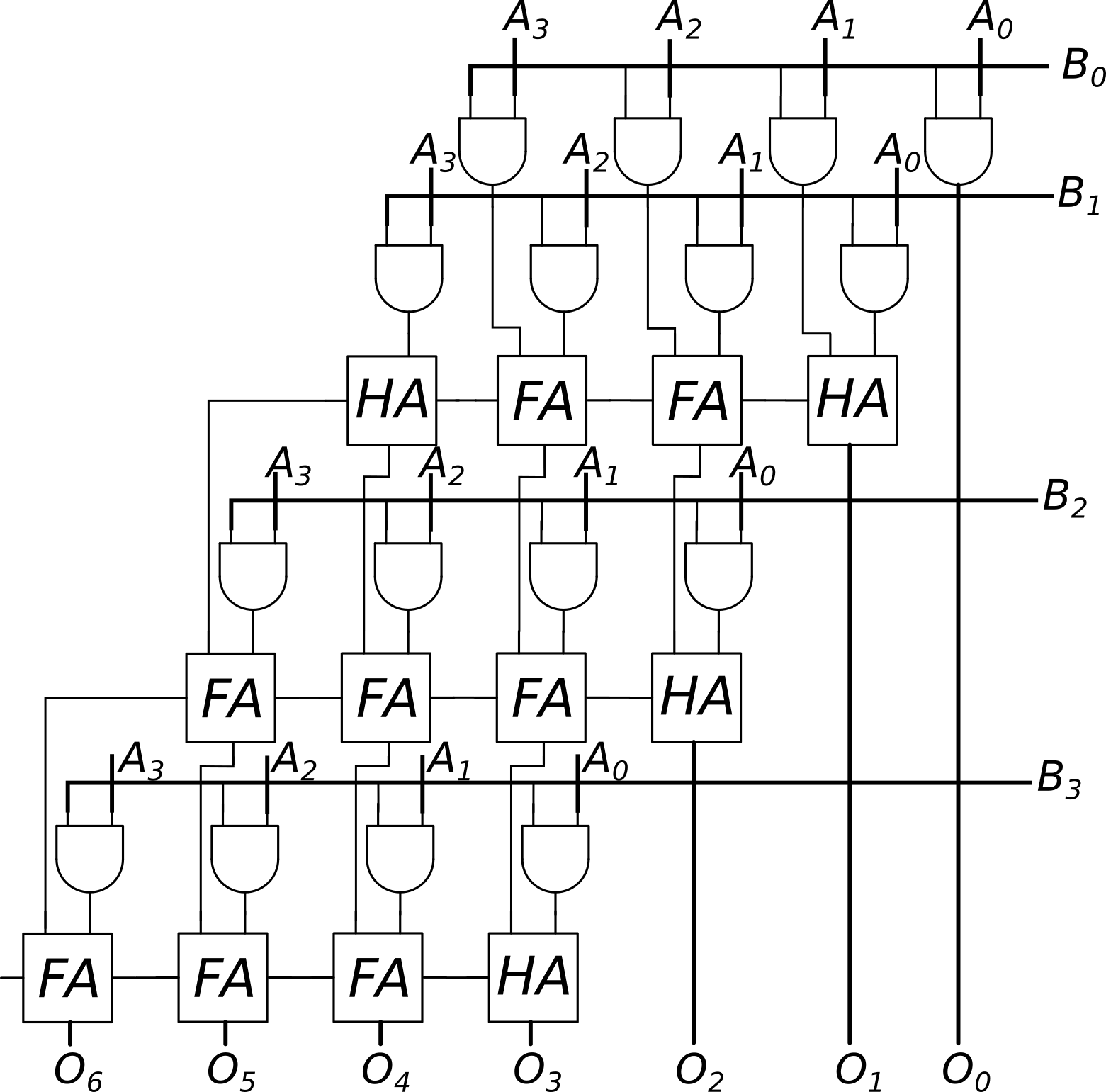}}
  \hfill
   \subfloat[Multiplier chip fabricated with STP2.]{\includegraphics[width=0.5\linewidth]{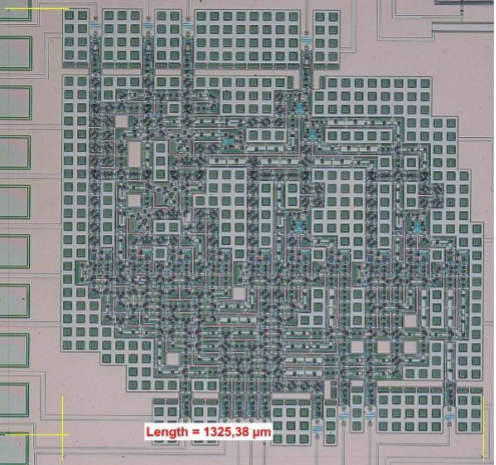}}
  \caption{Multiplier circuit fabricated for 4-bit multiplication. To reduce the number of delay JTLs, a DFF state is added to the multiplier, and it needs two clock cycles to generate a result.}
  \label{figure3ab}
\end{figure}

\begin{figure}[htpb]
\includegraphics[width=0.9\linewidth]{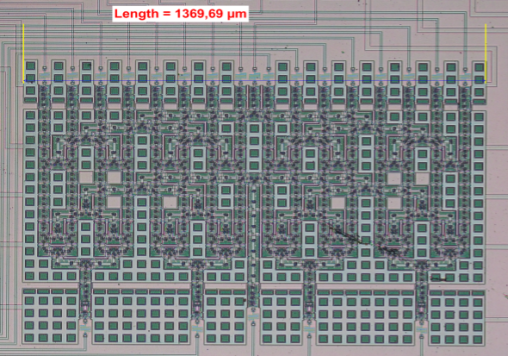}
\caption{4-bit 4 to 1 multiplexer circuit used in the input register. This unit matches the inputs' frequency to the high RSFQ frequency by four.}
\label{figure4}
\end{figure}

The circuit design for the ALU was implemented based on the parallel datapath, utilizing the CONNECT cell library \cite{ref28-1}. Subsequently, the fabricated circuits were produced at AIST CRAVITY (now known as QuFAB) using the STP2 process \cite{ref29-1}. Fig.~\ref{figure3ab} showcases that the 4-bit array multiplier circuit is one of the most complex circuit blocks within the ALU design. This particular circuit operates within two clock cycles, representing a significant component of the overall functionality of the ALU.
In Fig.~\ref{figure4}, we present the fabricated circuit for the buffer stage located at the input of the ALU. These buffers play a crucial role in receiving the outputs from the CMOS stage and loading them in conjunction with a signal from CMOS. The data is then serialized and pipelined, allowing for faster unloading to the input register of the ALU using a high-speed RSFQ clock. This configuration significantly enhances the communication bandwidth with the CMOS memory, facilitating an efficient and continuous data stream.

The fabrication process employed in this study incorporates 4 Nb layers, and the transmission lines also occupy a specific area within the circuit. However, if a denser process such as ADP and MIT LL SFQee5 \cite{ref30-1,ref31-1} with 9 Nb layers were utilized, the interconnects could be positioned beneath the active elements using passive transmission lines \cite{ref36}, resulting in a more compact, power-efficient and densely integrated circuit design.
\subsection{Measurement Setup}
\label{Sec-Test}
\begin{figure}[tpb]
\includegraphics[width=\linewidth]{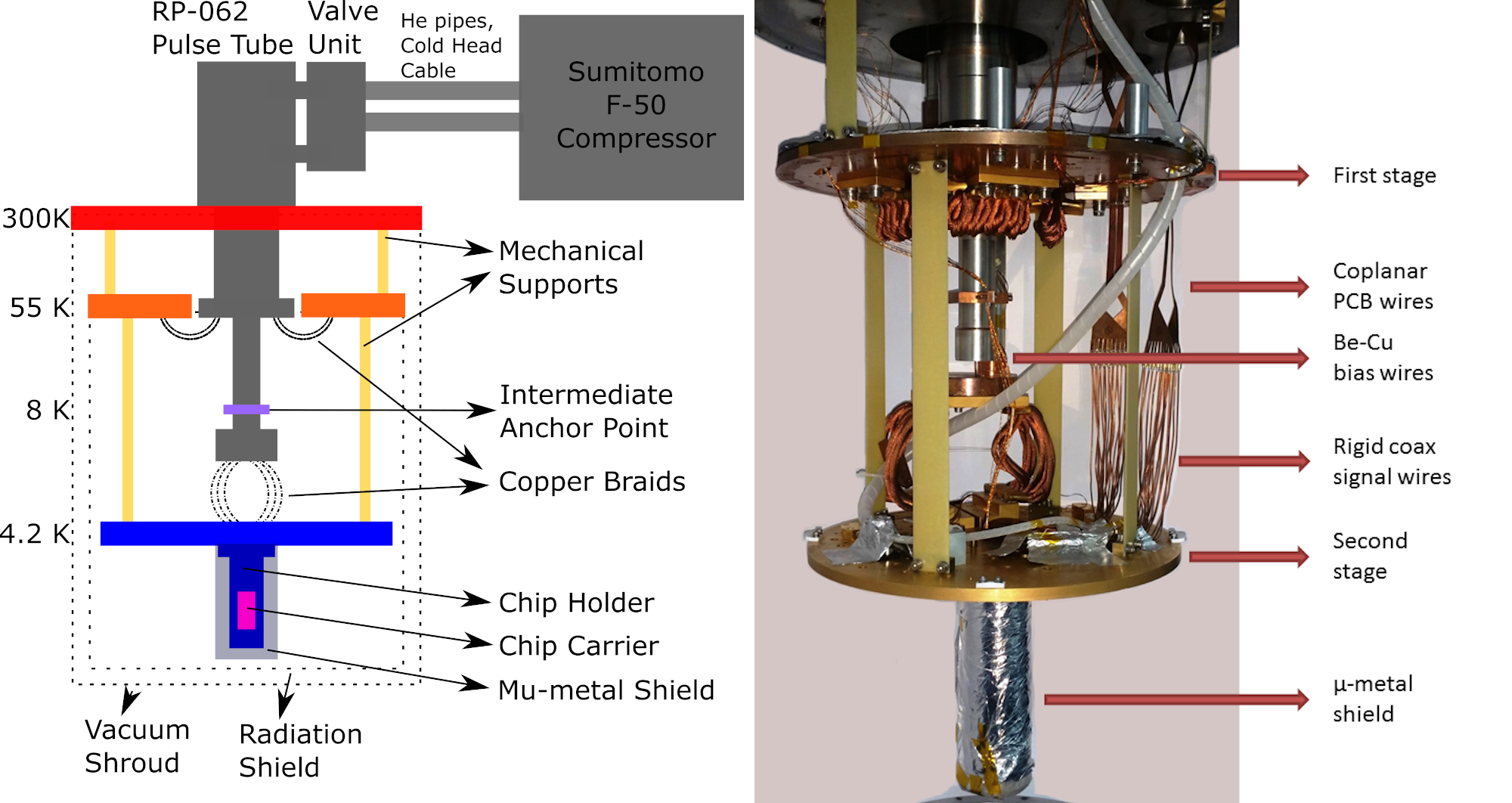}
\caption{The 4K Pulse-tube cryocooler used for the tests. The flat wiring provides good thermal isolation but limits the input frequency to 2.5GHz. RF and $\mu$-metal shields are essential for testing large RSFQ circuits.}
\label{figure5}
\end{figure}
In this work, we conducted the measurements using a custom-designed closed-cycle cryocooler system. We employed the Janis 4K pulse-tube Cryocooler with Sumitomo F-50 compressor, which offers a cooling power of 500 mW at 4.2 K. To minimize chip holder vibration, mechanical damping braids were installed at both the first and second stages, as depicted in Fig.~\ref{figure5}. These damping braids effectively reduced the vibration of the 4 K stage by one order of magnitude; however, they also decreased the cooling power of the second stage from 500 mW to 250 mW.

Operating relatively large circuits within a closed-cycle refrigerator system necessitates considering specific conditions. One significant challenge with such circuits is the current bias. Although power consumption for RSFQ circuits is typically much lower than the cooling capacities of the cryocooler due to the low operating voltage levels of 2.5 mV, they may require high current levels, leading to Joule heating in the wires. Good electrical conductors are usually good thermal conductors too. Hence the bias lines cannot have a very small resistance and must carefully consider the trade-offs between Joule heating and thermal heat load \cite{ref32}. The high current levels passing through the wires diminish the excess cooling power of the cryocooler, particularly in the second stage. We utilized an intermediate temperature point of approximately 8 K within the cold head to achieve thermal anchoring of the bias and signal lines. This approach reduces the thermal load on the 4.2 K stage compared to the direct connection to the 1st stage at 45 K. We refer to this anchor point as the middle stage.
\begin{figure}[htpb]
  \includegraphics[width=0.8\linewidth]{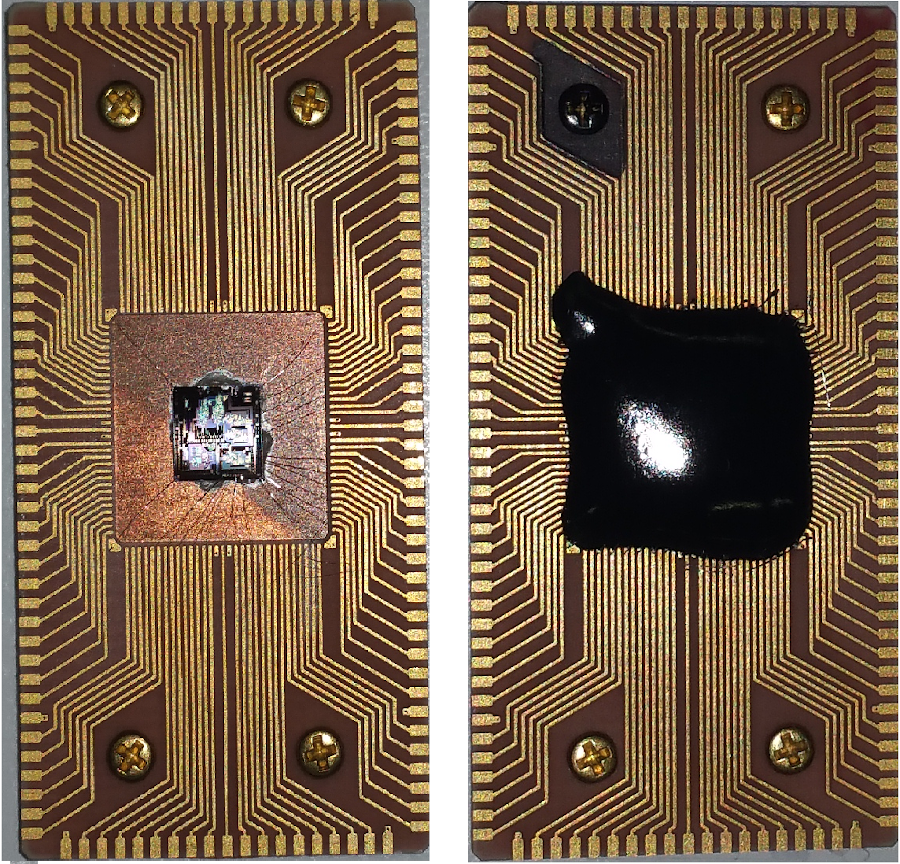}
  \caption{ALU chip packaging for implementation in the cold head of the cooler. The chip is placed and wire-bonded on the high-frequency PCB, and then we add Stycast epoxy for better contact between the PCB copper substrate and the chip surface.}
  \label{figure6}
\end{figure}
Fig.~\ref{figure5} illustrates the wiring and damping braids of the test system. Twisted Be-Cu wires were employed for DC bias lines, with a single wire resistance of $2.1 \Omega$ between room temperature and the chip holder. Approximately $0.6 \Omega$ exists within that value between the middle and second stages. Shielded co-planar Be-Cu flexible flat cables were utilized for high-frequency data transmission from the first to the second stage \cite{ref33}. Sem semi-rigid copper coaxial cables were implemented from the second stage to the chip. The heating occurring at the chip surface results in a temperature gradient between the chip and the cold head, affecting the superconductor's state and leading to chip heating despite significant power remaining at the second stage \cite{ref34}. A new packaging method, as shown in Fig.~\ref{figure6}, was employed for the ALU chip and its components to enhance power restrictions on the chip and accommodate large current surges. Detailed information regarding the system and packaging method can be found elsewhere \cite{ref34}.

Superconductor integrated circuits are susceptible to magnetic fields due to their reliance on Josephson Junctions. While external fields, such as the Earth's magnetic field, can be easily attenuated using high permeability shields, special attention must be given to the magnetic fields generated by the high current bias levels within the circuits. To mitigate the effect of the magnetic field resulting from bias currents, we employed the controlled ground return current method, which involves applying a negative value of the bias current in a nearby ground contact \cite{ref37}. However, one drawback of this method is that it increases the DC bias currents flowing through the bias lines, effectively doubling the applied circuit's current.

\subsection{Experimental Results}
We utilized a pattern generator system with a bandwidth of 600 MHz to generate the input signals and clock waveform for our experiments. We employed a precise multi-channel current source to provide the input bias currents for the circuits with low bias values. The output signals were sampled using a data acquisition card or logic analyzer after passing through a low-noise amplifier with a maximum gain of 60 dB. It is worth noting that the amplifiers limit the system's bandwidth at low frequencies. All instruments were controlled through a system PC to ensure minimal user interference during the tests.

\begin{figure}[htpb]
\includegraphics[width=0.9\linewidth]{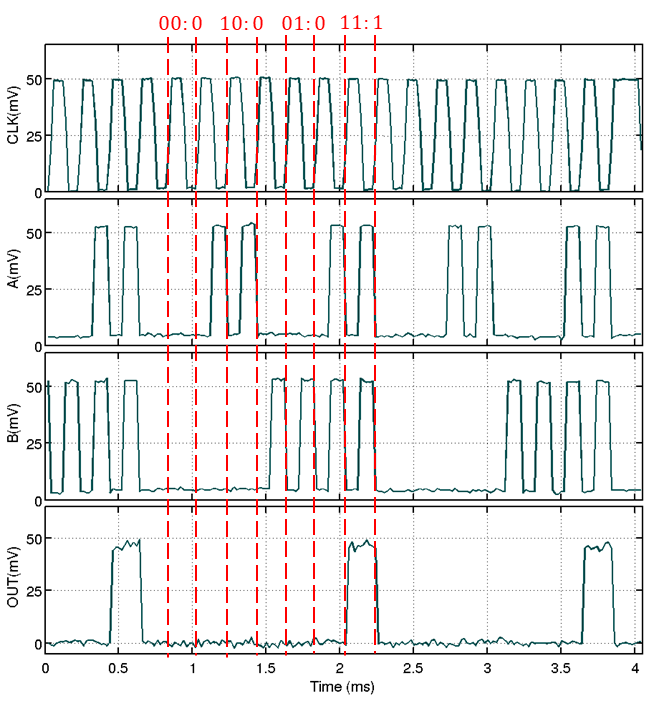}
\caption{Measurement results of the single J-AND gate. The input signal level to the DC/SFQ cell should be 1mA, provided by a 50mV source through a 50$\Omega$ resistor. The output is measured from the SFQ/DC cell, and the value is about 100$\mu V$ after 500$\times$ amplification becomes 50mV.}
\label{figure7}
\end{figure}

\begin{figure}[htpb]
  \subfloat[Multiplexer measurement]{\includegraphics[width=0.5\linewidth]{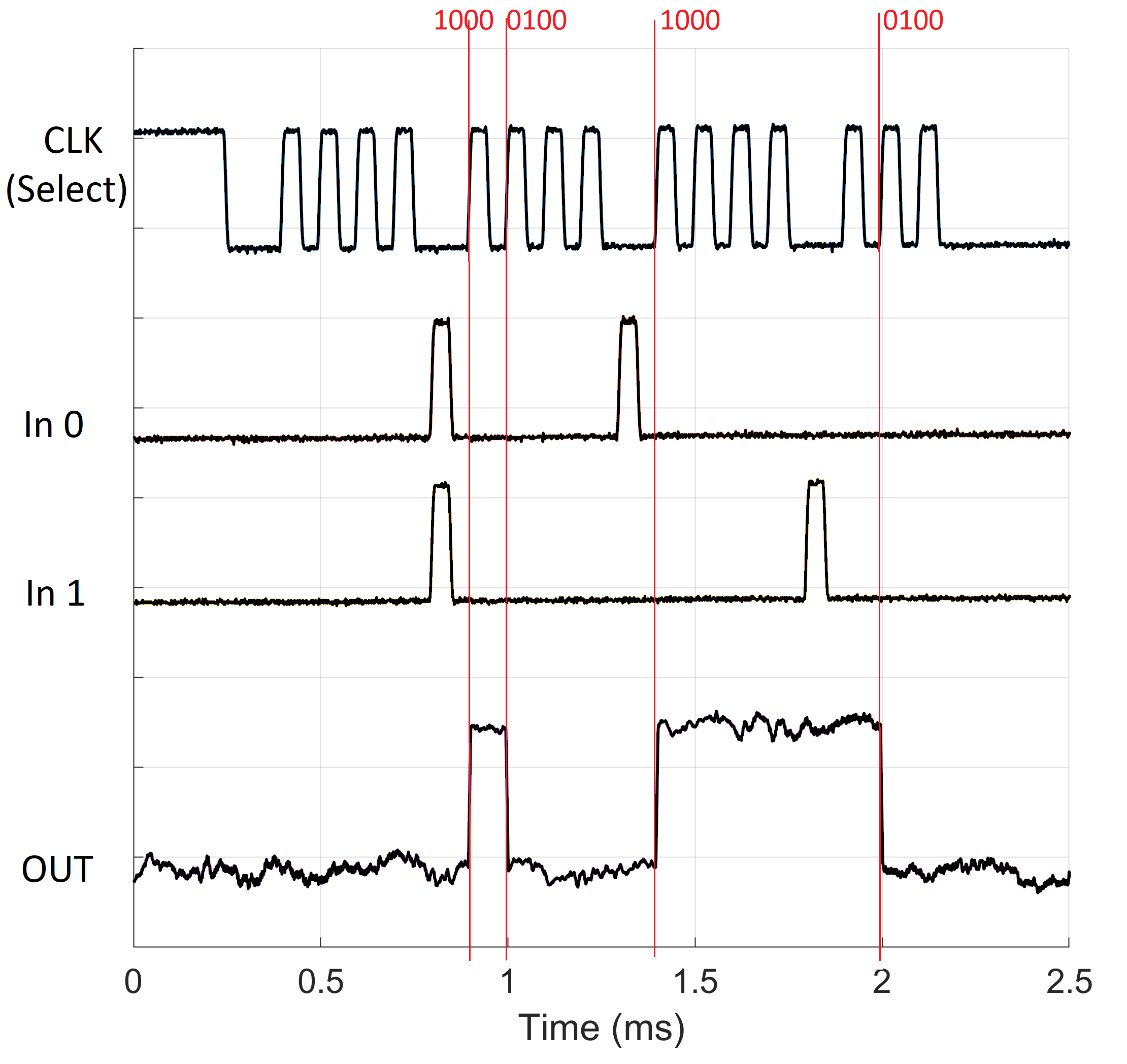}}
  \hfill
  \subfloat[Demultiplexer measurement]{\includegraphics[width=0.5\linewidth]{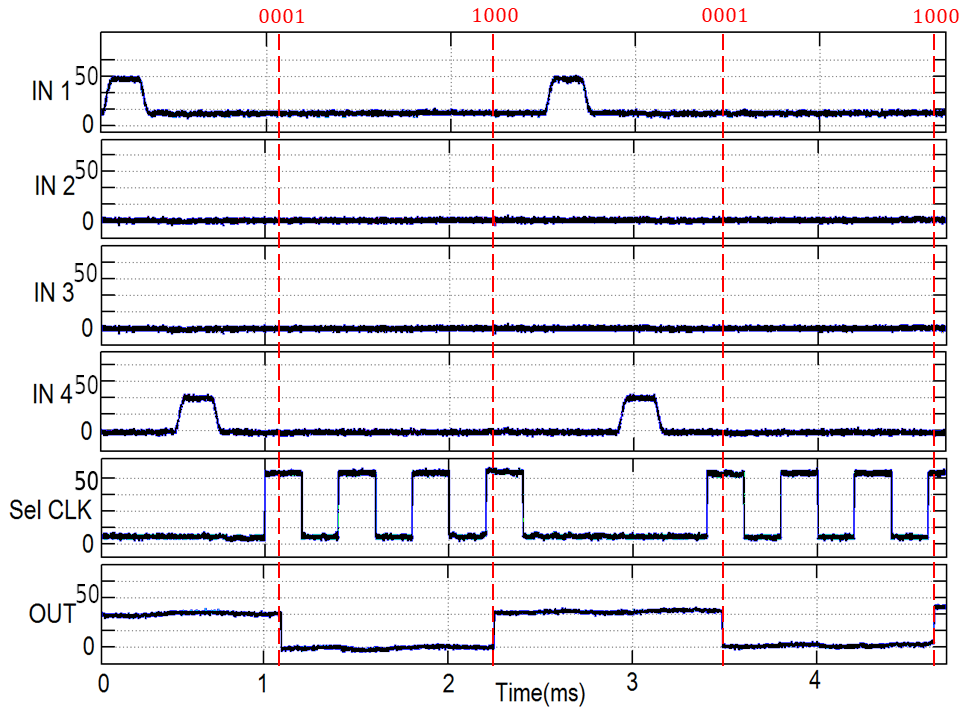}}
  \caption{Multiplexer and demultiplexer circuits' measurement results. (a) Multiplexer $2 \times 1$ circuit experimental results. Similar to the AND gate, the input values are 50mV, and the output is amplified to be 50mV. (b) The measurement result of the input register (demultiplexer $1 \times 4$).}
  \label{figure8}
\end{figure}

The functionality of the ALU sub-circuits was confirmed by applying various signal patterns at lower frequencies. The logic unit is designed to perform four functions: XOR, NOT, AND, and OR. The clock tree of the logic unit is distributed, ensuring that all delays from different cells are identical. Fig.~\ref{figure7} displays the measurement results for a simple Josephson AND logic gate in our test setup.
The multiplexer stage is crucial in determining the outputs based on the supplied selected bits. In our design, the instruction set signals were applied to multiplexers using clocked toggle flip-flop (T1) cells. The outputs of these cells were then fed into JAND cells, and their outputs were merged to generate the desired signal as the output. Fig.~\ref{figure8}(a) illustrates the result obtained for a toggle flip-flop-based multiplexer.

The input and output registers are crucial in communicating between the RSFQ ALU and the CMOS processor. Fig.~\ref{figure8}(b) illustrates the measurement results for the input register. As depicted in the figure, the four inputs from the CMOS stage are converted into four pulses synchronized with the correct clock cycles. The RSFQ ALU operates in a pipeline manner, ensuring that the outputs are processed in the same order as the inputs. This allows efficient data transfer and synchronization between the RSFQ ALU and the CMOS processor.
The adder circuit was initially designed using the Kogge-Stone (KS) architecture. However, due to the large size of the KS adder and the limited chip size, we had to change the design to a carry-look-ahead architecture. This modified design used clocked toggle flip-flop (TFF) cells as full adders. Additionally, to enable the subtract function, an XOR logic gate was introduced before the second input's register of the adder, and the two's complement of the second register was created.

\begin{figure}[htpb]
\centering
\includegraphics[width=0.9\linewidth]{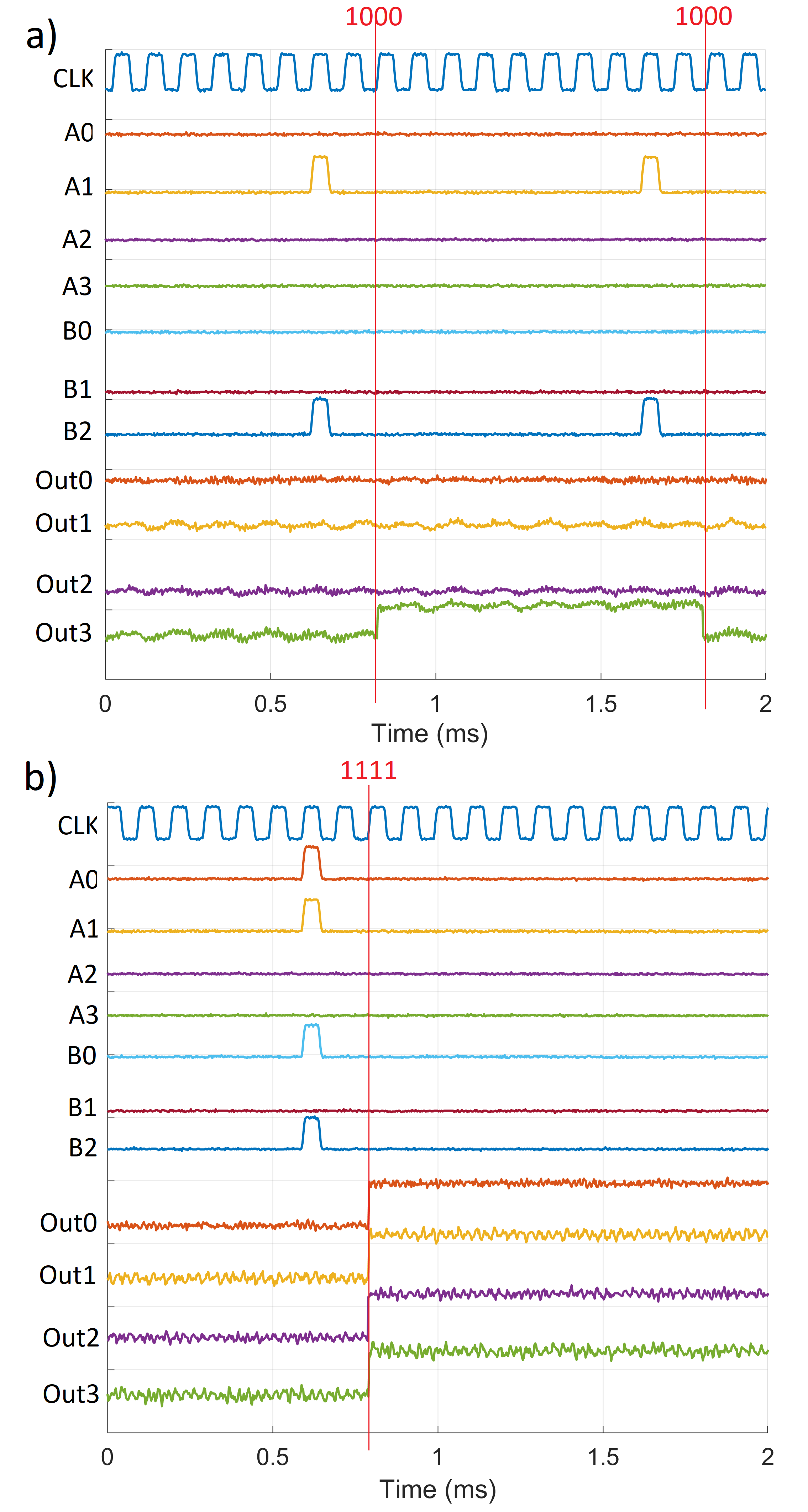}
\caption{The measurement result for the 4-bit multiplier cell. The voltages are acquired from a Textronix oscilloscope paired with a logic analyzer. a) 2$\times$4  and b) 3$\times$5. The inputs were applied from a pattern generator with variable probes set to 50mV. The outputs were measured on a Tektronic oscilloscope with four analog input channels for amplifier output and multiple digital input channels connected to the pattern generator to monitor the applied data. The multiplier needs two clock signals to generate output.}
\label{figure9}
\end{figure}

The multiplier circuit was also designed using full adders and could perform 4-bit multiplication in just two clock cycles. Fig.~\ref{figure9} displays the result obtained for the multiplier cell with two different input signals. It is worth noting that this multiplier circuit is one of the most complex circuits ever tested in a closed-cycle cryocooler, comprising over 2000 junctions and requiring a bias current of 350mA.
To assess the quality of the fabricated chip and measure the chip's surface temperature at different bias values, a series of DC-SQUIDs (Superconducting Quantum Interference Devices) were placed on the chip. Monitoring the I-V characteristic of these SQUIDs provides valuable insights into the chip's quality. Additionally, these measurements confirmed the presence of a temperature gradient between the cold head and the chip.

\begin{figure}[htpb]
\includegraphics[width=1\linewidth]{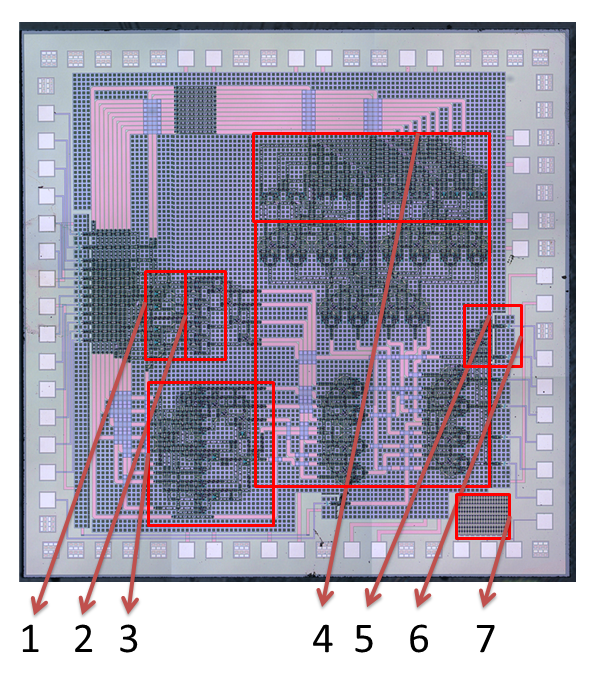}
\caption{The fabricated ALU with STP2 process \cite{ref38}. 1) The XOR gates at the input of the adder for subtraction, 2) Carry look-ahead adder stage, 3) Multiplier stage, 4) Logic Unit with NOT, AND, OR, and XOR gates, 5) The multiplexer stages, 6) Zero flag, 7) Series SQUIDs for temperature and fabrication parameter measurement.}
\label{figure10}
\end{figure}

After ensuring the correct operation of the individual components of the arithmetic logic unit (ALU), we combined them to form the complete circuit, as depicted in Fig.~\ref{figure10}. The ALU comprises over 9000 Josephson junctions (JJs) and requires approximately 950 mA of bias current. As stated before, to mitigate the susceptibility of RSFQ circuits to magnetic fields, we employed the controlled ground return current method \cite{ref37}. Considering the excess cooling power of 185mW available even after accounting for the Joule heating at the bias lines at 4.2 K, it is theoretically possible to maintain the ALU at 4.2 K with a bias current of 1900 mA, resulting in a total power consumption of 4.75 mW.
\section{Conclusion}\label{Sec4}
We have successfully designed, tested, and validated all the necessary components to develop a 4-bit arithmetic logic unit (ALU) with superconductor logic circuits. Apart from the arithmetic and logic circuits, we have also designed and tested interface sub-circuits to facilitate the synchronous operation of the RSFQ ALU with external CMOS circuits. The simulation results have exhibited a bias margin of approximately $\pm 18\%$ for the complete ALU, indicating a robust operational range. Moreover, a custom-made pulse-tube cryocooler has effectively demonstrated all the sub-circuits, showcasing satisfactory bias margins. In addition to the successful development and testing, we have fabricated a complete ALU, which will be further employed as a co-processor within a multi-chip-module setup. This integration will enhance processing capabilities and expand functionalities with other computational units.

\bmhead{Acknowledgments}

TUBITAK supported this work with project number 111E191.

The circuits were fabricated in the clean room for analog-digital superconductivity CRAVITY (now QuFAB) of the National Institute of Advanced Industrial Science and Technology (AIST) with the standard process 2 (STP2). 

The authors thank Prof. A. Fujimaki (Nagoya Univ., Japan) and his associates for kindly providing CONNECT cells.


\end{document}